\documentstyle[preprint,aps]{revtex}

\begin{document}

\tighten
\draft

\title{Weak measurement and the traversal time problem\protect{
\footnote{Talk given at the Adriatico Research Conference on 
``Tunnelling and its implications'', 30 July--2 August 1996, ICTP, Trieste}}}
\author{G. Iannaccone \cite{email}}
\address{Dipartimento di Ingegneria dell'Informazione,
Universit\`a degli Studi di Pisa, \\
Via Diotisalvi 2, I-56126 Pisa, Italy; \\
Institute for Microstructural Sciences, National
Research Council of Canada, \\ Ottawa, Canada K1A 0R6}
\maketitle

\begin{abstract}
The theory of weak measurement, proposed by 
Aharonov and coworkers, has been applied by Steinberg 
to the long-discussed traversal time problem. 
The uncertainty and ambiguity that characterize this concept
from the perspective of von Neumann measurement theory 
apparently vanish, and joint probabilities and conditional averages 
become meaningful concepts. 
We express the Larmor clock and some other well-known methods
in the weak measurement formalism.
We also propose a method to determine higher moments of the
traversal time distribution in terms of the outcome of a
gedanken experiment, by introducing an appropriate operator.
Since the weak measurement approach can sometimes lead to 
unphysical results, for example average negative reflection
times and higher moments, the interpretation of the results
obtained remains an open problem.
\end{abstract}

\section{Introduction}

In the last few years a new approach to measurement in 
quantum mechanics has been developed by
Aharonov and coworkers \cite{aharvaid90,aharanan93}. 
Their ``weak measurement'' approach differs
from the standard one (formalized by von Neumann\cite{neumann55}) in
that the interaction
between the measuring apparatus and the measured system is too weak 
to trigger a collapse of the wave function. 
Although an individual weak measurement of an observable has no meaning, 
one can obtain the expectation value to any desired accuracy
by averaging a sufficiently large number of such individual results. 

Avoiding wave function collapse allows the simultaneous
measurement of non-commuting observables (no violation of the 
uncertainty principle occurs because the individual measurements
of each observable are very imprecise). It also allows
a sound definition of  
conditional probabilities and their distribution: since the system 
evolves after the measurement as if unperturbed, it is possible 
to define averages of a  quantity
conditioned to a given final state of the system.
Moreover -- and this point is important if we are interested in the
duration of some process -- a typical
weak measurement is extended in time, i.e., the 
interaction between the meter and the system is not impulsive, but
has a finite duration.
As Steinberg has shown,\cite{steinber95_1,steinber95_2}
all these features make weak measurement theory a promising framework
for the study of traversal times in quantum systems, 
a problem that does not fit well 
within standard measurement theory.

In this paper, we show that the ambiguities which are present in
the formalism when
the traversal time problem
is studied with the tools of standard measurement
theory,\cite{brousala93} vanish
in the framework of the weak measurement approach. 
However, the {\it interpretation} of the 
weak measurement results remains open.
The outline of the paper is as follows: In Section 2 we present briefly
the weak measurement theory (WMT), in a ``minimalistic" way, i.e., 
concentrating on only those aspects of WMT that are directly relevant 
to the traversal time problem.
We apply the technique to this problem
in Sec. 3 and in Sec. 4 show  
that several well known methods for defining and
calculating average traversal times are particular realizations
of the weak measurement approach. In Sec. 5 we go further and 
introduce an operator for the time spent in a region of space
in an attempt to obtain higher moments of the 
traversal and dwell time distributions. A short discussion of
open problems ends the paper.

\section{Weak measurement: a ``minimalist'' formulation}

In this section we describe the generic 
{\em gedanken} experiment and compare the standard measurement theory
of von Neumann with the weak measurement theory of Aharonov and
coworkers. For the scope of this paper we do not need to
push the theory as far as Aharonov et al.\cite{aharanan93}
and will limit the discussion to weak measurements 
on an ensemble of systems, staying clear of the more controversial
issues of weak measurements on a single system and the reality of the wave 
function (i.e., the possibility of measuring the wave function of 
a single system). We use a minimalist approach to weak
measurement theory treating it as a potentially useful extension of
standard measurement theory, based on a ``weak'' system-apparatus
interaction Hamiltonian.

The experimental setup consists of a system $\Sigma$ and a
measuring device $M$ evolving -- when isolated -- under the Hamiltonians
$\hat{H}_\Sigma$ and $\hat{H}_M$, respectively.
Let $q$ be the canonical variable of the meter that we use as a pointer, 
and let $\pi$ be its conjugate momentum. The corresponding
operators are $\hat{q}$ and $\hat{\pi}$ with $[\hat{\pi},\hat{q}] =
-i\hbar$.

To measure an observable $\hat{A}$ of the system $\Sigma$, 
let the system and apparatus interact through the Hamiltonian
\begin{equation}
\hat{H}_{int} = g(t)\hat{\pi}\hat{A}(t)
\label{hamint}
,\end{equation}
where $g(t) = Gh(t)$, $G$ is a constant
and $\int_{-\infty}^{+\infty} h(t)dt = 1$. 
Let $h(t)$ be non-zero only for $t \in (t_i,t_f)$. 

The system $\Sigma$ and the meter $M$ evolve 
independently with Hamiltonian $\hat{H}_0 = \hat{H}_\Sigma + \hat{H}_M$
until time $t_i$,
then undergo  the interaction governed by 
$\hat{H}_{int}$, and, after time $t_f$,
continue their evolution under $\hat{H}_0$.  What is
measured is the position of the meter at time $t_f$. 

Let us denote by $|\psi_0(t)\rangle$, $|\phi_0(t) \rangle$,
and
$|\Phi_0(t) \rangle \equiv |\psi_0(t) \rangle \otimes |\phi_0(t) \rangle$
the states representing
the system $\Sigma$, the meter $M$, and their combination
$\Sigma$ plus $M$, respectively, evolving without
mutual interaction,
and by $|\Phi(t)\rangle$ the state of the combined system
after the switching on of the interaction  $\hat{H}_{int}$
at time $t_i$.
Since the system $\Sigma$ and the meter $M$ do not interact
before 
time $t_i$, $|\Phi(t)\rangle = |\Phi_0(t)\rangle$ for $t <
t_i$.

For simplicity, we will consider $\hat{H}_M = 0$, that is the state of
the meter is static until the interaction is turned on,
so that we can use
$|\phi_i \rangle = | \phi_0(t_i) \rangle$ for
the state of the meter before time $t_i$.
Moreover, after the interaction is switched off, at $t_f$, the
state of the meter in each component of the linearly superposed
entangled state no longer changes with time.

In the Schr\"odinger picture,\cite{sakurai85}
\begin{equation}
|\Phi(t_f)\rangle = \hat{U}(t_f,t_i) | \Phi(t_i)\rangle
\label{Psievolv}
,\end{equation}
where $\hat{U}(t_f,t_i)$ is the evolution operator
\begin{equation}
\hat{U}(t_f,t_i) = \left( \exp \left\{ -\frac{i}{\hbar} \int_{t_i}^{t_f} 
                        [ \hat{H}_0(t) + \hat{H}_{int}(t) ] dt
                        \right\} \right)_+
,\label{evol_op}
\end{equation}
and the $+$-subscript denotes time ordering of the integrals in the
terms of the Taylor series expansion of the exponential function.
In the following, we will indicate a state in the
Heisenberg representation by omitting its dependence on time:
for instance, $| \Phi \rangle$ is the state $| \Phi(t) \rangle$ in the
Heisenberg representation, and is obtained as
$| \Phi \rangle = \hat{U}(t_f,t) \Phi(t)\rangle$.

\subsection{Standard Measurement}

In the von Neumann procedure, \cite{neumann55}
$t_f$ tends to $t_i$, i.e, $h(t) \approx
\delta(t-t_f)$, and what is measured is the value of the observable $A$
at the instant of time $t_f$.

In the time interval $(t_i,t_f)$, $\hat{H}_{int}$ is the 
dominant term in the Hamiltonian and, 
from (\ref{Psievolv}) and (\ref{evol_op}), we  have
\begin{equation}
|\Phi(t_f)\rangle 
\approx e^{-\frac{i}{\hbar}G\hat{\pi}\hat{A}(t_f)} |\Phi(t_i)\rangle
\label{psif_strong}
.\end{equation}

The probability density of pointer position $q$ after the interaction is
\begin{equation}
f(q) \equiv \langle \Phi | q \rangle \langle q | \Phi \rangle
= \sum_n \langle \Phi | a_n,q \rangle \langle a_n,q | \Phi \rangle
\label{fdqstrong}
,\end{equation}
where $\{ | a_n(t) \rangle \}$ is a complete 
set of eigenstates of $\hat{A}(t)$.
Straightforward calculation\cite{neumann55} yields
\begin{equation}
f(q) = \sum_n |c_n(t_f)|^2 | \phi_i(q-Ga_n)|^2
\label{fdqstrongfinal} 
,\end{equation}
where $c_n(t) \equiv \langle a_n(t) | \psi_0(t) \rangle$ and
$ \phi_i(q-Ga_n) = \langle q - Ga_n | \phi_i \rangle$.

It is worth noticing that if the initial pointer position $q$ is 
precisely defined, that is $|\phi_i(q)|^2 \approx \delta(q)$, 
the probability density of the final position is a sum of quasi-delta
functions in one-to-one correspondence with each of the eigenvalues 
of $\hat{A}$.

\subsubsection{Distribution of the pointer position}

The first two moments of the pointer position distribution are now easy
to obtain. If we take an initial distribution of $q$ centered at $q=0$,
the mean value of $q$ at time $t_f$ is
\begin{equation}
\langle q \rangle_f \equiv
\langle \Phi| \hat{q} | \Phi \rangle
= \int q f(q) dq = G \langle A(t_f) \rangle
,\end{equation}
and the mean square value of $q$ is
\begin{equation}
\langle \hat{q}^2 \rangle_f \equiv 
\langle \Phi |q^2|\Phi \rangle = \int q^2 f(q) dq
= \langle \hat{q}^2 \rangle_i + G^2 \langle A^2(t_f) \rangle
,\label{qsquare}
\end{equation}
so that 
\begin{equation}
( \Delta q_f)^2 
\equiv 
\langle q^2 \rangle_f - ( \langle q \rangle_f )^2
=
(\Delta q_i)^2 + G^2 (\Delta A_f)^2
,\end{equation}
where $\Delta q_f$, $\Delta q_i$, and $\Delta A_f$ are the
standard deviations of final and initial pointer positions,
and of the observable $A$ at time $t_f$, respectively.
The integrals without explicit limits are from $-\infty$ to $+\infty$.

\subsubsection{Verification of the unperturbed state}

It is interesting to calculate the probability that the state
of the system $\Sigma$ under observation is not changed. In order
to do so, we calculate the probability $P_0$ of verification of the
unperturbed state $|\psi_0\rangle$ at time $t_f$, i.e. 
\begin{equation}
P_0(t_f) \equiv \langle \Phi | \psi_0 \rangle \langle \psi_0 | \Phi
\rangle
= 
\int \langle \Phi | \psi_0,q \rangle 
\langle \psi_0,q | \Phi\rangle dq
.\end{equation}
If we remember that $|\psi_0(t) \rangle = \sum_n c_n(t) | a_n(t)\rangle$
we obtain
\begin{equation}
P_0(t_f) = \sum_{n, m} |c_n(t_f)|^2 |c_m(t_f)|^2 
\int \phi_i^*(q-G a_n) \phi_i(q-Ga_m) dq
\label{p0strong}
,\end{equation}
but if $\Delta q_i \ll G \Delta a$, where $\Delta a$ is the
minimum difference between the eigenvalues of $\hat{A}$ 
($\Delta a = \min_{n\neq m} \{ | a_n - a_m | \}$),
the integral in (\ref{p0strong}) is practically zero
when $n \neq m$, so that we can write
\begin{equation}
P_0(t_f) \approx \sum_n |c_n(t_f)|^4 \leq \max_{n} \{ |c_n(t_f)|^2 \}
\label{p0strongfin}
.\end{equation}
Equation (\ref{p0strongfin}) shows that the initial state is
conserved only if it is an eigenstate of $\hat{A}$; if this
is not the case, the evolution of the system is strongly
affected by the measurement. 
As will be shown in the next section, this problem does not exist
in the weak measurement approach, due to the fact that
the evolution of the system is 
perturbed only to order $o(G)$ (by $o(G)$ we mean a term such that
$\lim_{G \rightarrow 0} o(G)/G = 0$).

\subsection{Weak measurement}

Weak measurement is characterized by the fact that 
the Hamiltonian for the interaction $\hat{H}_{int}$ is
small enough to be considered as a small perturbation of
the Hamiltonian $\hat{H}_0$ of the isolated system $\Sigma$,
and the initial uncertainty in the position of the
pointer $q$ is much greater than $G$ times the maximum
separation between different eigenvalues of $\hat{A}$.

Most importantly, the interaction does not have to be 
impulsive, but can have a finite duration of time.
This additional flexibility is a 
great advantage for measurements made over finite
intervals of time.

According to perturbation theory, \cite{baym} we can write
\begin{equation}
| \Phi(t_f) \rangle  = | \Phi_0(t_f) \rangle 
- \frac{i}{\hbar} \int_{t_i}^{t_f} \hat{U}_0(t_f,t)        
					\hat{H}_{int}(t)| \Phi(t) \rangle dt
\label{phipert}
,\end{equation}
where
\begin{equation}
\hat{U}_0(t_f,t) = \left( \exp \left\{-\frac{i}{\hbar} 
\int_{t}^{t_f} \hat{H}_0(t')dt' \right\} \right)_+
\end{equation}
is the evolution operator of the isolated system $\Sigma$.
First order approximation on (\ref{phipert}) gives
\begin{equation}
| \Phi(t_f) \rangle =(1+\hat{o}(G))| \Phi_0(t_f) \rangle 
		- \frac{i}{\hbar} G \hat{\pi} \int_{t_i}^{t_f} 
					\hat{U}_0(t_f,t)
					\hat{A}(t)| \Phi_0(t) \rangle h(t) dt
,\label{phipert2}
\end{equation}
where $\hat{o}(G)$ indicates a generic operator whose averages are $o(G)$.

If we introduce the hermitian operator in the Heisenberg picture
\begin{equation}
{\cal I}_H(\hat{A}) \equiv
\int_{t_i}^{t_f} \hat{U}_0(t_f,t)\hat{A}(t)\hat{U}_0^*(t_f,t) h(t) dt
,\label{inta}
\end{equation}
we can write ({\ref{phipert2}) as
\begin{equation}
|\Phi\rangle = [ 1 - \frac{i}{\hbar} G \hat{\pi} 
		{\cal I}_H(\hat{A}) +
		\hat{o}(G) ] | \Phi_0 \rangle
\label{phipertfin}
.\end{equation}
Now we define $A_w$, the {\em weak value} of the operator $\hat{A}$, as 
\begin{equation}
A_w \equiv \langle \Phi_0 | {\cal I}_H(\hat{A})|\Phi_0 \rangle
= \int_{t_i}^{t_f} h(t) \langle \psi_0(t) | \hat{A}(t) | \psi_0(t)
\rangle dt
\label{aweak}
\end{equation}
The probability density of $q$ after time $t_f$ is
$f(q) \equiv \langle \Phi |q \rangle \langle q | \Phi \rangle$
and can be written, by using (\ref{phipertfin}) and (\ref{aweak}), as 
\begin{eqnarray}
f(q) & = & 
\langle \Phi_0 | 1 + \frac{i}{\hbar}G \hat{\pi}
                {\cal I}_H(\hat{A}) + \hat{o}(G)|q \rangle 
\langle q | 1 - \frac{i}{\hbar} G \hat{\pi}
                {\cal I}_H(\hat{A}) + \hat{o}(G)| \Phi_0 \rangle 
,\nonumber \\
& = &
\langle \Phi_0 | \exp ( \frac{i}{\hbar}G A_w \hat{\pi} )
                + \hat{o}(G)|q \rangle 
\langle q | \exp ( - \frac{i}{\hbar} G A_w \hat{\pi} )
                + \hat{o}(G)| \Phi_0 \rangle \nonumber \\
& = & | \phi_i(q-GA_w)|^2 + o(G)
\label{fdqweak}
.\end{eqnarray}
Except for terms of $o(G)$, the final distribution of pointer positions
is equal to the initial one translated by $G$ times the weak value of
$\hat{A}$.
It is worth noticing that if the interaction is impulsive (i.e.,
$h(t) \approx \delta(t-t_f)$), we have 
$A_w \approx \langle A(t_f) \rangle $.

\subsubsection{Distribution of pointer position}

The mean pointer position and the variance are, from (\ref{phipertfin})
and (\ref{fdqweak}),
respectively
\begin{equation}
\langle q \rangle_f \equiv \langle \Phi | \hat{q} | \Phi \rangle 
 = \int q f(q) dq = G A_w + o(G)
\label{weakavg}
\end{equation}
and
\begin{equation}
( \Delta q_f )^2 
= \langle q^2 \rangle_f - ( \langle q \rangle_f)^2
= (\Delta q_i)^2 +o(G)
.\label{weakstddev}
\end{equation}
The average pointer position gives us the weak value of $A$; on the
other hand, the variance does not give us additional information,
because the weak measurement is very imprecise, due to the fact that
the initial pointer distribution is very broad and the interaction
is weak. Averaging over many identical experiments gives the right
mean value, but does not tell us anything about the dispersion of
the observed quantity, which is completely swamped by the
dispersion in pointer position.

\subsubsection{Verification of the unperturbed state}

A fundamental property of a WM is that the evolution of $\Sigma$
is practically not perturbed.
In fact,
verification of the initial state, using (\ref{phipertfin}), yields
\begin{eqnarray}
P_0(t_f) = \langle \Phi | \psi_0 \rangle \langle \psi_0 | \Phi \rangle 
 = 1 + o(G)
\label{poweak}
\end{eqnarray}
This means that several weak measurements of different observables
on a single system can be performed. 
As a general property, and therefore even for non commuting observables, 
the order of successive measurements is not important.

\subsubsection{Conditional averages}

While conditional averages are not well defined within
standard measurement theory\cite{brousala93}, they can be introduced
in an
unambiguous way within WMT, as a consequence of
eq. (\ref{poweak}) discussed above.
Suppose that we want to measure the average of $\hat{A}$ conditioned
to the verification of a given final state which is assumed,
without loss of generality, to be a member $|\chi_n \rangle$
of an orthonormal basis $\{ | \chi_n \rangle \}$, for $n=1 \dots N$,
of the Hilbert space of $\Sigma$.
Since $|\chi_n \rangle \langle \chi_n|$ and $\hat{q}$ commute,
we can perform a standard measurement of both of them
when the interaction is over, i.e., after time $t_f$.
Then, we keep only the
readings of $q$ corresponding to a positive verification of
$|\chi_n \rangle$, and calculate the ``conditional''
probability distribution of the
collected readings $f(q)^{(n)}$, which is of the form 
\begin{equation}
f(q)^{(n)} = \frac{
\langle \Phi | \chi_n, q \rangle \langle \chi_n, q | \Phi \rangle }
{ \langle \Phi | \chi_n \rangle \langle \chi_n | \Phi \rangle}
= |\phi_i(q - G A_w^{(n)})|^2 + o(G)
,\end{equation}
where
\begin{equation}
A_w^{(n)} \equiv \frac{\langle \chi_n| {\cal I}_H(\hat{A})|\psi_0 \rangle}
			{\langle \chi_n | \psi_0 \rangle} = 
\frac{1}{\langle \chi_n | \psi_0 \rangle}
\int_{t_i}^{t_f} \langle \chi_n(t) | \hat{A}(t) | \psi_0(t) \rangle h(t) dt
\label{aweakpost}
.\end{equation}

$A_w^{(n)}$ is the weak value of $\hat{A}$ for a system which is 
postselected in the state $| \chi_n \rangle$ (and preselected in
the state $| \psi_0 \rangle$). 
To order $o(G)$, the probability amplitude distribution of the
meter's pointer is equal to the initial one translated by a 
quantity proportional to $A_w^{(n)}$.
When defining $A_w$ in (\ref{aweak})
we did not specify a post-selected state; actually, to not perform a
post-selection is equivalent to post-selecting the state $|\psi_0 \rangle$
because, as we showed in (\ref{poweak}), verification of $|\psi_0 \rangle$
is positive with probabililty very close to one. Therefore, 
$A_w^{(n)}$ of (\ref{aweakpost}) is equal to $A_w$
of (\ref{aweak}) to order $o(G)$ if $| \psi_0 \rangle$
happens to be $| \chi_n \rangle$.
If it does not, then we can write
$ | \psi_0 \rangle = \sum_n p_n | \chi_n \rangle$,
where $p_n = \langle \chi_n | \psi_0 \rangle$,
and have
\begin{equation}
A_w = \sum_n |p_n|^2 A_w^{(n)}
\label{weakdecompose}
.\end{equation}
It is important to notice that, while $A_w$ is always real, $A_w^{(n)}$ is
in general complex valued.

From (\ref{qsquare}) and (\ref{aweak}) we find that
the conditional average and the standard deviation of the 
pointer position are, respectively:
\begin{equation}
\langle q \rangle_f^{(n)} \equiv
\int_{-\infty}^{+\infty} f(q)^{(n)} q dq = 
G {\rm Re} \{ A_w^{(n)} \} + o(G) 
\end{equation} 
and       
\begin{equation} 
( \Delta q_f^{(n)} )^2 = 
\langle q^2 \rangle_f^{(n)} - ( \langle q \rangle_f^{(n)} )^2 =
( \Delta q_i )^2 + o(G)
,\end{equation}
independent of $n$.

In addition, from (\ref{weakavg}) and (\ref{weakdecompose}) we have
\begin{equation}
\langle q \rangle_f 
=
\sum_n | p_n |^2  
\langle q \rangle_f^{(n)}
,\label{condprob}
\end{equation}
that is, the well known  sum law of conditional probabilities holds true
for pointer position readings.

\section{Weak measurement and traversal times}

Measurement of the time duration of some process requires that 
the observed system and the meter interact for a finite time,
a situation for which the concept of weak measurement 
seems to be particularly well suited.
Moreover, as we have just seen, WMT could also allow us to 
calculate conditional averages of a given temporal quantity
for various outcomes of the unperturbed system.

A well known and widely accepted result in the field of tunneling
times is the dwell time, i.e. the average time spent by a particle in
the region $\Omega$ irrespective of its final state. \cite{but_larmor83} 
If $|\psi_0 \rangle$ is the state describing the
particle, the dwell time in the interval $(t_i,t_f)$ 
is postulated to be\cite{jawoward88}
\begin{equation}
\langle t_D \rangle =
\tau_D(t_i,t_f) = \int_{t_i}^{t_f} \langle \psi_0(t) | \hat{P}_\Omega |
\psi_0(t) \rangle dt
\label{dwelltime}
,\end{equation}
where $\hat{P}_{\Omega}$ is the projection operator on the region $\Omega$.
As can be seen, (\ref{dwelltime}) is the mean value of $\hat{P}_\Omega$
integrated over $(t_i,t_f)$. It is hard to imagine this time as a
result of a standard measurement, because $\hat{P}_\Omega$ is not a 
quantum non demolition (QND) variable\cite{peres93} and,
if $t_1 \neq t_2$, then $\hat{P}_\Omega(t_1)$
and $\hat{P}_\Omega(t_2)$ do not commute. 

However, (\ref{dwelltime}) can be obtained as a result of a weak
measurement.
In fact, if we take $\hat{A} = \hat{P}_\Omega$, and $h(t)$ as constant in
$(t_i,t_f)$, the interaction Hamiltonian is
\begin{equation}
\hat{H}_{int} = G h(t) \hat{\pi} \hat{P}_\Omega
\label{hint}
,\end{equation}
and from (\ref{aweak}) we have
\begin{equation}
P_{\Omega w} = 
\frac{1}{t_f - t_i} \int_{t_i}^{t_f} 
\langle \psi_0(t) | \hat{P}_\Omega | \psi_0(t) \rangle dt
.\label{pomegaweak}
\end{equation}
Combining (\ref{dwelltime}) and (\ref{pomegaweak}) yields
the dwell time as
\begin{equation}
\langle t_D \rangle = \tau_D(t_i,t_f) = (t_f -t_i) P_{\Omega w}
= \lim_{G \rightarrow 0} 
\frac{\langle q \rangle_f (t_f -t_i)}{G}
,\label{dwelltimeread}
\end{equation}
where we have used the fact that 
$\langle q \rangle_f = G P_{\Omega w} +o(G)$.

Suppose we are interested in the mean time spent in $\Omega$ for some
specified final state of the particle.
Decomposition of dwell times in terms
of particles evolving to a final state $|\chi_n \rangle$ is problematic
within standard measurement theory, as has been pointed out many 
times: \cite{brousala93}
the difficulty is that projection onto a region $\Omega$  and
projection onto a final state $| \chi_n \rangle$ involve non commuting
operators, and there are no rules uniquely
specifying how to build operators for
quantities involving non commuting operators (this is also the 
reason for conditional probabilities being problematic).

The ambiguity vanishes within the weak measurement approach: the
weak value of $\hat{P}_\Omega$ for a system postselected in the
final state $| \chi_n \rangle$ is, according to (\ref{aweakpost}),
\begin{equation}
P_{\Omega w }^{(n)} =
\frac{1}{\langle \chi_n | \psi_0 \rangle}
\frac{1}{t_f -t_i}
\int_{t_i}^{t_f} 
\langle \chi_n(t_f) | \hat{P}_\Omega | \psi_0(t_f) \rangle dt
\label{pomegawn}
\end{equation}
Therefore, the average time spent in $\Omega$ from time $t_i$ to $t_f$
by a particle starting in the state $|\psi_0 \rangle$ and finally found
in the state $| \chi_n \rangle$ is
\begin{equation}
\langle t_D \rangle^{(n)} = \tau_D^{(n)} 
\equiv \frac{(t_f -t_i)\langle q \rangle_f^{(n)}}{G}
= (t_f -t_i) {\rm Re}\{ P_{\Omega w}^{(n)} \}
.\label{dwelltimereadpost}
\end{equation}
Summation over different final states holds: given
$ | \psi_0 \rangle = \sum_n p_n | \chi_n \rangle$
then, dropping the dependence on the time interval, we can write,
from (\ref{condprob}), (\ref{dwelltimeread}), and
(\ref{dwelltimereadpost}),
\begin{equation}
\langle t_D \rangle = \sum_n |p_n|^2 \langle t_D \rangle ^{(n)} 
.\end{equation}

\section{Weak measurement and well known methods for obtaining traversal
times}
 
In this section we want to demonstrate that some well known approaches to
the calculation of tunneling times can be seen as
particular examples of weak measurement, 
each corresponding to a different measuring apparatus.

In particular, we will focus our attention on methods
based on the Larmor clock, \cite{but_larmor83,rybachen67,baz_larmor67} 
on Feynman path-integrals, \cite{sokobask87,iannpell94}, and 
on absorption probabilities. \cite{golufelb90} 
All of these procedures are based on the
application of a small perturbation (a magnetic field, a real potential,
an imaginary potential, respectively) to the region of interest.
After that, the state of the particle evolves in time, and
we attempt to extract the information about the time spent
in the region of interest from some aspect of the
perturbed wave function (the spin, the phase, or the
amplitude, respectively depending on the kind of perturbation
applied). In order not to perturb the evolution of the state too much,
we let the perturbation tend to zero \cite{iannpell94}. It has been
demonstrated \cite{muga,iannpell_larmor94} that all the ``probes'' mentioned
above lead to the same result.

Let us now write two formulas that will be very useful in the 
remainder of this section.
From Appendix A, the weak value of an operator $\hat{A}$
for a system postselected in the state $| \chi_n \rangle$, defined
in (\ref{aweakpost}), can be written as
\begin{equation}
A_w^{(n)} = 
\left.
\frac{\partial}{\partial G}
\frac{\langle \chi_n(t_f),\pi|\hat{q}|\Phi\rangle}
        {\langle \chi_n,\pi|\Phi_0\rangle}
\right|_{\pi, G = 0}
=
\left.
\frac{
\langle \chi_n, \pi | i \hbar \frac{\partial}{\partial (G\pi)}
| \Phi \rangle }
{\langle \chi_n, \pi | \Phi_0 \rangle }
\right|_{G\pi = 0}
.\label{awutile2}
\end{equation}
where the second equality is true if
$\hat{q}$ can be written as $\hat{q} = i \hbar \partial/\partial \pi$
in the $\pi$-representation and $| \Phi \rangle$ 
depends only upon the product
$G\pi$ [as it obviously does for the interaction Hamiltonian
(\ref{hamint})].

\subsection{Real constant potential}

Let us start with a constant real potential applied only in $\Omega$
and only for $t_i < t < t_f$: the perturbation Hamiltonian is 
$\hat{H}_{int} = \hat{H}_{V} = f(t) V \hat{P}_\Omega$,
with $f(t) = 1$ for $t \in (t_i,t_f)$ and zero otherwise. \cite{iannpell94} 
In order to translate this perturbation into the formalism of
weak measurement, we can write $V$ in the $\pi$-representation
as $V= G \pi /(t_f - t_i)$. Now the perturbative potential
acting on the system $\Sigma$ is of the form (\ref{hint}).

In this case, the weak value of the operator $\hat{P}_\Omega$ for
a system postselected in the state $| \chi_n \rangle$ is,
according to (\ref{awutile2}),
\begin{equation}
P_{\Omega w}^{(n)} = 
\left.
\frac{
\int
\langle \chi_n | {\bf r} \rangle
\langle {\bf r},\pi | i \hbar \frac{\partial}{\partial (G\pi)}| \Phi \rangle
d^3 {\bf r}
}{
\int
\langle \chi_n | {\bf r} \rangle
\langle {\bf r},\pi | \Phi_0 \rangle d^3{\bf r}
}
\right|_{G\pi=0}
.\label{pomegaweakpost}
\end{equation}
We use the convention of omitting the limits of integration when
the integrals run over the whole space.
Given that $V$ is proportional to $\pi$, we can
write $\Phi({\bf r}, V) = \langle {\bf r}, \pi| \Phi \rangle$
and $\chi_n({\bf r}) = \langle {\bf r} | \chi_n \rangle$,
so that (\ref{dwelltimereadpost}) becomes
\begin{equation}
\langle t_D \rangle^{(n)} = 
(t_f -t_i) {\rm Re}\{P_{\Omega w}^{(n)}\} =
{\rm Re} \left\{
\left.
\frac{ 
\int \chi_n^*({\bf r}) i \hbar 
\frac{\partial}{\partial V} 
\Phi({\bf r},V) d^3{\bf r}
}{
\int \chi_n^*({\bf r}) \Phi_0({\bf r}, V) d^3{\bf r}
}
\right|_{V=0}
\right\}
\label{tdwellv}
\end{equation}

Note that (\ref{tdwellv}) is exactly the expression for the
average time spent by a particle
in the region $\Omega$ obtained by using the Feynman path-integral
technique. \cite{sokobask87} If the final state is $|{\bf r} \rangle$,
i.e., the state corresponding to 
a particle found to be at ${\bf r}$ at time $t_f$, the weak
value of the average time is then
\begin{equation}
\langle t_D \rangle^{({\bf r})} = 
	{\rm Re} \left. \left\{ \frac{i \hbar}{ \Phi({\bf r},V)} 
	\frac{\partial \Phi({\bf r},V)}{\partial V} \right\}
	\right|_{V=0}
.\end{equation}
which is exactly the same expression obtained for the stay time
defined in [18].

\subsection{Pure imaginary potential}

A pure imaginary potential is often used in optics to
simulate the absorption of photons by a material. What 
happens in this case is that the probability density of the
particle is not conserved, because it decreases exponentially
in $\Omega$, with a time constant proportional to the applied
imaginary potential. The information about the average time
spent in $\Omega$ by the particle is therefore obtained by 
calculating how much of the total probability has been absorbed.

The perturbation Hamiltonian in this case is \cite{iannpell_larmor94}
\begin{equation}
\hat{H}_{int} =
\hat{H}_{\Gamma} = - f(t) \frac{i \Gamma}{2} \hat{P}_{\Omega}
\end{equation}
which is of the form (\ref{hint}) if we put
$ \Gamma = 2 i G \pi/(t_f - t_i)$.
Analogously to (\ref{pomegaweakpost}) and (\ref{tdwellv}) we have
\begin{equation}
\langle t_D \rangle^{(n)} = \tau_D^{(n)} = -
\left.
\frac{\int \chi_n^* ({\bf r}) 2 \hbar 
\frac{\partial}{\partial \Gamma}
\Phi({\bf r},\Gamma) d^3{\bf r}
}{
\int \chi_n^*({\bf r}) \Phi_0({\bf r},\Gamma) d^3 {\bf r}
}
\right|_{\Gamma = 0}
,\end{equation}
where we have put $\Phi({\bf r}, \Gamma) = 
\langle {\bf r}, \pi | \Phi \rangle$. This result, again,
corresponds to the one obtained in [21].

\subsection{Magnetic Field}

The well known Larmor clock method \cite{rybachen67,baz_larmor67}
involves applying an infinitesimal magnetic field in the $z$-direction,
confined to the region $\Omega$.
The spin, which is initially polarized in the $x$-direction,
precesses in the $x$-$y$ plane with the Larmor frequency
$\omega_L = e B/m$ when the spin is ``in'' $\Omega$.
The spin polarization in the $y$-direction plays the role
of pointer position. Let us consider as the
perturbation Hamiltonian only the component which acts on the
spin of the particle \cite{iannpell_larmor94}
\begin{equation}
\hat{H}_{int} =
\hat{H}_{B} = f(t) \frac{\hbar \omega_L}{2} \hat{\sigma}_z
		\hat{P}_\Omega
\label{hamperb}
,\end{equation}
where $\hat{\sigma}_x$, $\hat{\sigma}_y$, and $\hat{\sigma}_z$ are
the Pauli spin matrix operators.
In this case $\hat{\pi} = \hbar \hat{\sigma}_z/2$ acts as the pointer
momentum and we put
$G = \omega_L ( t_f -t_i )$, so that 
(\ref{hamperb}) takes the form (\ref{hint}).

We have $\hat{\sigma}_x | \psi_0 \rangle = | \psi_0 \rangle$
because the initial state of the system is an eigenstate 
of $\hat{\sigma}_x$.
From
\begin{equation}
[ \hat{\sigma}_y, \frac{\hbar}{2} \hat{\sigma}_z ] | \psi_0 \rangle
=
i \hbar \hat{\sigma}_x | \psi_0 \rangle = i \hbar | \psi_0 \rangle
\end{equation}
it immediately follows that $\hat{q} = \hat{\sigma}_y $
and $\hat{\pi} = \hbar \hat{\sigma}_z/2$ are the appropriate conjugate
pointer operators. With this choice  (\ref{awutile2}) becomes
\begin{equation}
P_{\Omega w}^{(n)} = 
\left.
\frac{\partial}{\partial G}
\frac{\langle \chi_n, \pi | \hat{\sigma}_y | \Phi \rangle}
	{\langle \chi_n, \pi | \Phi_0 \rangle}
\right|_{\pi,G = 0}
\end{equation}
and
\begin{equation}
\langle t_D \rangle ^{(n)} = \tau_D^{(n)} = {\rm Re}
	\left\{ 
	\left.
	\frac{\partial}{\partial \omega_L}
	\frac{\langle \chi_n, \pi | \hat{\sigma}_y | \Phi \rangle}
		{\langle \chi_n, \pi | \Phi_0 \rangle}
	\right|_{\omega_L,\pi = 0}
	\right\}
.\label{timepostlarmor}
\end{equation}
As  is easy to see by comparison with (18) of Ref. [21],
expression (\ref{timepostlarmor})
for the time spent in $\Omega$ is equal to the result obtained by 
Rybachenko \cite{rybachen67} and Baz'. \cite{baz_larmor67} 

\section{Higher moments of time distributions}

As is clear from (\ref{weakstddev}) weak measurements are not useful
for obtaining higher moments of a distribution for the time 
spent in $\Omega$. In fact, the spread of final positions of the
pointer is equal to the initial one to $o(G)$. The only
way within WMT of obtaining, say, the $l$th order moment of an operator
$\hat{A}$, is to build a meter sensitive to $\hat{A}^l$. This
should have an interaction Hamiltonian of the form 
$\hat{H}_{int}^{[l]} =G h(t) \hat{\pi}_l \hat{A}^l(t)$. 
In principle, there is no
fundamental problem with this, and several meters can act 
simultaneously on the same system. 

The crucial point is that we need to use an operator for the time
spent in $\Omega$, and not just the projector over $\Omega$ as
we did in section 3.
In this section we will use the ``sojourn time'' operator previously
introduced by Jaworski and Wardlaw.\cite{jawoward92}
It is
consistent with the results
of section 3 and 4, 
and is easy to obtain
from the definition of mean dwell time (\ref{dwelltime}).

\subsection{An operator for the time spent in $\Omega$}

In the Heisenberg representation, the dwell time defined by
(\ref{dwelltime}) can be written as
\begin{equation}
\langle t_D \rangle = \tau_D(t_i,t_f) = 
\langle \psi_0 | \hat{t}_{\Omega H} | \psi_0 \rangle 
\label{dwellop}
\end{equation}
if we just define
\begin{equation}
\hat{t}_{\Omega H} \equiv 
\int_{t_i}^{t_f} 
\hat{U}_0(t_f,t') \hat{P}_\Omega \hat{U}^*_0(t_f,t') dt'
= (t_f - t_i) {\cal I}_H(\hat{P}_\Omega)
,\label{tomegaop}
\end{equation}

In the Schr\"odinger representation, the operator $\hat{t}_\Omega$
corresponding to $\hat{t}_{\Omega H}$, is
\begin{equation}
\hat{t}_{\Omega}(t) =
\hat{U}_0^*(t_f,t) \hat{t}_{\Omega H} \hat{U}_0(t_f,t)
.\label{tomegaopsch}
\end{equation}
For a gedanken experiment with a meter sensitive to $\hat{t}_{\Omega}$,
the interaction Hamiltonian is 
\begin{equation}
\hat{H}_{int}^{[1]} = G_1 h(t) \hat{\pi}_1 \hat{t}_{\Omega}(t)
,\label{hamint1}
\end{equation}
where $h(t) = (t_f - t_i)^{-1}$ for $t \in (t_i,t_f)$, and $0$ otherwise;
$\hat{\pi}_1$ and $\hat{q}_1$ are the conjugate momentum and
position of the meter's pointer.
From (\ref{tomegaopsch}) if follows that
${\cal I}_H(\hat{t}_{\Omega})$ defined by (\ref{inta})
is equal to $\hat{t}_{\Omega H}$.
Application of (\ref{aweak}) and (\ref{aweakpost})
then leads to 
\begin{equation}
t_{\Omega w}  = \langle \psi_0 | {\cal I}_H(\hat{t}_{\Omega}) 
			| \psi_0 \rangle 
 = \langle \psi_0 | \hat{t}_{\Omega H} | \psi_0 \rangle
,\hspace{1cm} 
t_{\Omega w}^{(n)} =
\frac{\langle \chi_n | \hat{t}_{\Omega H} | \psi_0 \rangle}
{\langle \chi_n | \psi_0 \rangle}
.\label{tauhweakpost}
\end{equation}

If we take $\langle t_D \rangle $ defined in (\ref{dwelltime}), and 
$\langle t_D^{(n)} \rangle$
defined in (\ref{dwelltimereadpost}), we can write
\begin{equation}
\langle t_D \rangle = \lim_{G_1 \rightarrow 0} 
	\frac{\langle q_1 \rangle_f}{G_1} = t_{\Omega w} 
, \hspace{1cm}
\langle t_D \rangle ^{(n)} = 
\lim_{G_1 \rightarrow 0} 
\frac{\langle q_1 \rangle_f^{(n)}}{G_1} = 
{\rm Re} \{ t_{\Omega w}^{(n)} \}
.\end{equation}
As  can be seen, $\hat{t}_\Omega$ leads to the same result
as $\hat{P}_\Omega$, in the measurement of average traversal times.

\subsection{Higher moments}

By the means of $\hat{t}_\Omega$, we can measure any moment of order $l$
of the distributions of times spent in $\Omega$. We need to use
a meter whose corresponding interaction Hamiltonian is of the kind
\begin{equation}
\hat{H}_{int}^{[l]} =
G_l h(t) \hat{\pi}_l \hat{t}_\Omega^l
,\label{hamintl}
\end{equation}
where $\hat{\pi}_l$ and $\hat{q}_l$ are the operators corresponding to
the conjugate momentum and position of the meter's pointer.
The average of the $l$th power of the time spent in $\Omega$ by a particle
finally found in the state $| \chi_n \rangle$ is
\begin{equation}
\langle t^l_D \rangle^{(n)} 
\equiv
\lim_{G_l \rightarrow 0} 
\frac{\langle q_l \rangle_f^{(n)}}{G_l} 
=
{\rm Re}\{ \langle t^l_{\Omega} \rangle_w^{(n)} \}
,\label{taudln}
\end{equation}
with
\begin{equation}
\langle t_\Omega^l \rangle_w^{(n)} 
\equiv
\frac{\langle \chi_n | ( \hat{t}_{\Omega H} )^l | \psi_0 \rangle}
{\langle \chi_n | \psi_0 \rangle}
.\label{tauomegalpost}
\end{equation}
Only those pointer position readings
corresponding to a postselected state $| \chi_n \rangle$ are averaged.
It is worth noticing that the sum rule of conditional averages
is satisfied, i.e., if 
$|\psi_0 \rangle = \sum_n p_n | \chi_n\rangle$, 
then, for any integer $l$, 
\begin{equation}
\langle t_D^l \rangle = \sum_n |p_n|^2 \langle t_D^l \rangle^{(n)}
\end{equation}
It is also important to point out, while 
$\langle t_D^l \rangle$ is positively defined,
the conditional averages $\langle t_D^l \rangle^{(n)}$ are not.
The lack of this important property has to prevent us from
interpreting
these quantities as the moments of a distribution of
actual times spent by the electron in the
region $\Omega$.

\subsection{Comparison with some results in the literature}

The second moment of $t_D$, according to (\ref{taudln})
and (\ref{tauomegalpost}), is
$\langle t_D^2 \rangle = 
\langle \psi_0 | t_{\Omega H}^2 | \psi_0 \rangle$;
if we remember that 
$\hat{t}_{\Omega H} = {\cal I}_H(\hat{P}_{\Omega})(t_f -t_i)$,
we obtain 
\begin{eqnarray}
\langle t_D^2 \rangle & = & 
(t_f - t_i)^2
\langle \psi_0 | {\cal I}^2_H(\hat{P}_\Omega)| \psi_0 \rangle
\nonumber \\
& = & (t_f - t_i)^2
\int d^3r \langle \psi_0 | {\cal I}_H(\hat{P}_\Omega) | {\bf r}
\rangle \langle {\bf r} | {\cal I}_H(\hat{P}_\Omega) |\psi_0 \rangle
\nonumber \\
& = & \int d^3r |t_\Omega^{({\bf r})}|^2 |\psi_0({\bf r}, t_f)|^2
\label{td2}
\end{eqnarray}
where, as can be easily obtained from (\ref{pomegawn}) and
(\ref{dwelltimereadpost}), 
$t_\Omega^{({\bf r})}$ is the weak value of the time spent
in $\Omega$ by a particle finally found in $\bf r$.

Eq. (\ref{td2}) is essentially equal to the result obtained for the
second moment of the dwell time by a few works based on the 
path-integral approach.
\cite{leavaers90,sokoconn91,iannpell94,schuziol89}  

We would also point out that
the second moment of the time spent in $\Omega$ for a particle which
is post-selected in position $\bf r$ at time $t_f$, i.e., 
\begin{equation}
\langle t_D^2 \rangle^{(\bf r)}
= 
{\em Re}\left\{ 
\frac{ 
\langle {\bf r} | \hat{t}^2_{\Omega H} | \psi_0 \rangle}
{ \langle {\bf r}| \psi_0 \rangle }
\right\}
= \frac{
\langle \psi_0 | \hat{P}_{\bf r} \hat{t}^2_{\Omega H}
                 + \hat{t}^2_{\Omega H} \hat{P}_{\bf r} | \psi_0 \rangle}
{ \langle \psi_0 | \hat{P}_{\bf r} | \psi_0 \rangle }
,\end{equation}
where $P_{\bf r} = |{\bf r} \rangle \langle {\bf r} |$
is in general different from the time proposed
in Ref. [18] on the basis of the path integral approach,
that, in this formalism, would be equal to
$t_\Omega^{2\,({\bf r})} = 
\langle \psi_0 | \hat{t}_{\Omega H} \hat{P}_{\bf r} 
\hat{t}_{\Omega H} | \psi_0
\rangle /\langle \psi_0 | \hat{P}_{\bf  r} | \psi_0 \rangle$

\subsection{Relation between higher moments and the measurement of the
first moment}

In this section we show
that the higher order moments of $t_D$ obtained in Sec. 5.2
can be obtained also from the wave function  $|\Phi \rangle $
of the system plus meter perturbed by the Hamiltonian for the first
moment $\hat{H}_{\rm int} = G_1 h(t) \hat{\pi}_1 \hat{t}_\Omega(t)$.
In fact, if we multiply both numerator and denominator of
(\ref{tauomegalpost}) by $\langle \pi | \phi_i \rangle$, 
and substitute (\ref{ultima}) in the numerator, we obtain
\begin{equation}
\langle t^l_\Omega \rangle_w^{(n)}
=
\frac{1}{\langle \chi_n,\pi | \Phi_0 \rangle}
\left.
\langle \chi_n, \pi | \left( \frac{i \hbar}{\pi_1}
\frac{\partial}{\partial G_1} \right)^l 
| \Phi \rangle 
\right|_{G_1=0}
.\end{equation}

If we put $\hat{\lambda} = \hat{\pi}_1 G_1$, so that $\hat{H}_{\rm int} =
\hat{\lambda} h(t) \hat{t}_\Omega(t)$, and call
$\Phi(\lambda,{\bf r},t_f) 
= \langle \pi_1, {\bf r} | \Phi(t_f) \rangle$
we can write for any integer $l$
\begin{equation}
\langle t_D^l \rangle^{({\bf r})}
= {\rm Re} 
\left\{ \frac{1}{\Phi_0(\lambda,{\bf r}, t_f)}
\left( i \hbar \frac{\partial}{\partial \lambda} \right)^l
\Phi(\lambda,{\bf r},t_f) 
.
\label{quasiquasi}
\right\}
\end{equation}

Let us just point out that, while the form of (\ref{quasiquasi})
is exactly equal to the $l$-th complex moment of the dwell
time distribution obtained on the basis of path integrals
\cite{leavaers90,sokoconn91}, the meaning is substantially different, 
since the perturbative Hamiltonian
used in path-integral approaches is of the kind
$\hat{H}_{\rm pi} = \hat{\lambda} h(t) \hat{P}_\Omega(t_f -t_i)$,
while the perturbative Hamiltonian used for
obtaining (\ref{quasiquasi}) is $H_{\rm int}^{[1]}$ given by
(\ref{hamint1}). It is clear, for example,
that the former is local in space, while the latter is not.

\section{Discussion}

Steinberg \cite{steinber95_1,steinber95_2} has argued
that weak measurement theory is a promising tool
for the study of the traversal time problem. Its major advantages over the
standard measurement theory are the flexibility to treat
interactions between a system and a measuring apparatus 
that are extended in time, 
and the possibilty of defining conditional averages for events
corresponding to non commuting operators. 
Both these properties are due to the fact that a weak measurement 
prevents the wave function of the system from collapsing.

We have shown that within WMT not only mean dwell
and traversal times but also the averages of any higher
powers of the time spent by particles in a region $\Omega$,
conditioned to any final state of the system, can be
mathematically defined in terms of the outcome of
gedanken experiments.

Unfortunately, there are severe problems of physical interpretation.
As already pointed out for the special cases of the 
Larmor\cite{iannpell94,leavaers_dwell89} and Salecker-Wigner clocks
\cite{rickrosssal}, WMT may predict negative results for the average
time spent by reflected
particles on the far side of a barrier.
In addition, as shown here, the conditional averages of any
power of the time spent in $\Omega$ are not positively defined
within WMT. 
These unphysical results prevent us from interpreting them in terms of
actual time spent by particles in the spatial region $\Omega$. 

To remain on firm ground, we are compelled to
consider them as just quantities with the dimensions of time
describing the response
of a degree of freedom $q$ of an apparatus to an
interaction with particles that is constant in
time over a finite time interval, linear, and proportional
to a particle's presence
in $\Omega$. Clearly, further investigation is required to
learn whether these quantities can be fruitfully used
to describe the time-dependent behaviour of $\Sigma$ itself,
i.e., apart from the particular interaction with the meter.

\section{Acknowledgments}

The author would like to thank W. R. McKinnon,
and B. Pellegrini, and especially C. R. Leavens 
for many stimulating discussions and comments on the manuscript.
This work has been supported by the Italian 
Ministry of University and Scientific and Technological Research,
by the Italian National Research Council (CNR), and by the
National Research Council (NRC) of Canada.

\appendix

\section{Derivation of (36)}

We can start  from the Eq. (\ref{aweakpost}), where $A_w^{(n)}$
is defined. If we multiply both numerator and denominator by
$\langle \pi | \phi_i \rangle$ for $\pi = 0$ we have
\begin{equation}
A_w^{(n)} =
\left.
\frac{
\langle \chi_n, \pi | {\cal I}_H{(\hat{A})} | \Phi_0 \rangle
}{
\langle \chi_n, \pi | \Phi_0 \rangle
}
\right|_{\pi = 0}
.\label{aweakapp}
\end{equation}
Now, we have just to remember that, $\hat{1} = [\hat{q} , \hat{\pi}]/i\hbar$ 
and to substitute this formula into (\ref{aweakapp}) in order to
obtain
\begin{equation}
A_w^{(n)} =  
\left.
\frac{1}{i \hbar} 
\left[
\frac{
\langle \chi_n, \pi | \hat{q} \hat{\pi} {\cal I}_H{(\hat{A})} | \Phi_0 \rangle
}{
\langle \chi_n, \pi | \Phi_0 \rangle}
 - \pi  
\frac{
\langle \chi_n, \pi | \hat{q} {\cal I}_H{(\hat{A})} | \Phi_0 \rangle
}{
\langle \chi_n, \pi | \Phi_0 \rangle
}
\right]
\right|_{\pi = 0}
;\label{aweakapp2}
\end{equation}
the second term of this expression vanishes for $\pi = 0$.
If we substitute (\ref{derivg1}) for $l=1$ into
the first term to the right of (\ref{aweakapp2}),
we obtain Eq. (\ref{awutile2}).

\section{A few formulas from perturbation theory}

Let $| \Phi_I(t) \rangle$ and $\hat{H}_{int}^{(I)}(t)$ be
the system wave function and the interaction Hamiltonian,
respectively, in the interaction representation, \cite{baym} i.e.,
\begin{eqnarray}
| \Phi_I(t) \rangle & \equiv & \hat{U}_0(t_f,t) | \Phi(t) \rangle
,\label{phiint} \\
\hat{H}_{int}^{(I)}(t) & \equiv &  
\hat{U}_0(t_f,t) \hat{H}_{int}(t) \hat{U}^*_0(t_f,t)
,\label{hamintint}
\end{eqnarray} 
where $\hat{U}_0(t_f,t) = \exp \{ -i/\hbar \int_t^{t_f} \hat{H}_0(t')
dt' \} $ 
is the evolution operator.

From (\ref{phiint}) we have that $| \Phi_I(t_f) \rangle =
| \Phi(t_f)\rangle = | \Phi \rangle$ and
$| \Phi_I(t_i) \rangle = | \Phi_0(t_f)\rangle = | \Phi_0 \rangle$, 
therefore 
\begin{equation}
|\Phi \rangle =
\left( 
\exp \left\{ 
\frac{1}{i \hbar} 
\int_{t_i}^{t_f} \hat{H}_{int}^{(I)}(t) dt \right\} \right)_+ 
| \Phi_0 \rangle
.\label{phipertexp}
\end{equation}
where the $+$-subscript denotes time-ordering.

If we take $\hat{H}_{int} = G h(t) \hat{\pi} \hat{A}(t) $
as given by (\ref{hamint}), with $h(t) = (t_f-t_i)^{-1}$ for
$t \in (t_i,t_f)$ and zero otherwise,
and put it in (\ref{hamintint}) and (\ref{phipertexp}), we obtain
\begin{equation}
| \Phi \rangle = \left( \exp \left\{ \frac{G}{i \hbar} \hat{\pi}
{\cal I}_H(\hat{A}) \right\} \right)_+| \Phi_0 \rangle
.\label{phipertexpa}
\end{equation}
Writing the exponential in (\ref{phipertexp}) as a sum yields
\begin{equation}
| \Phi \rangle =
\sum_{m=0}^{\infty} 
\frac{1}{m!}
\left( \frac{G}{i \hbar} \right)^m \hat{\pi}^m 
\left( [ {\cal I}_H(\hat{A}) ]^m\right)_+
| \Phi_0 \rangle
,\label{phipertpow}
\end{equation}
from which we obtain
\begin{equation}
\left. \frac{\partial ^l}{\partial G^l} | \Phi \rangle \right|_{G=0}
=
\frac{1}{(i \hbar)^l} \hat{\pi}^l \left( [{\cal I}_H(\hat{A}) ]^m
\right)_+
| \Phi_0 \rangle
,\label{derivg1}
\end{equation}

If we choose $\hat{A}(t)=\hat{t}_\Omega(t)$, we have the
additional advantage that
${\cal I}_H(\hat{t}_\Omega) = \hat{t}_{\Omega H}$ does
not depend on time, so that time-ordering does not matter,
and we can write
\begin{equation}
\left. \frac{\partial ^l}{\partial G^l} | \Phi \rangle \right|_{G=0}
=
\frac{1}{(i \hbar)^l} \hat{\pi}^l \hat{t}_{\Omega H}^l
| \Phi_0 \rangle
,\label{derivg}
\end{equation}
from which we have, after projection onto the state
$| \chi_n, \pi \rangle$,
\begin{equation}
\left. \langle \chi_n, \pi | \hat{t}^l_{\Omega H} | \Phi_0 \rangle
=
\langle \chi_n, \pi | \left( \frac{i \hbar}{\pi} 
                             \frac{\partial}{\partial G}
		      \right)^l
| \Phi \rangle \right|_{G=0}
.\label{ultima}
\end{equation}


\end{document}